\documentclass{svmultm}

\authorrunning{F. Cabarle, H. Adorna, M. Mart\'inez--del--Amor}
\titlerunning{Simulating Spiking Neural P systems without delays using GPUs}

\usepackage{amsmath,amssymb,latexsym,graphicx,url,listings, algorithm, algorithmic}

\usepackage{todonotes,ulem}

\begin{document}

\title*{Simulating Spiking Neural P systems without delays using GPUs } \author{Francis Cabarle\inst{1}, Henry Adorna\inst{1}, Miguel A. Mart\'inez--del--Amor\inst{2}}
\institute{
Algorithms \& Complexity Lab\\
Department of Computer Science\\
University of the Philippines Diliman\\
Diliman 1101 Quezon City, Philippines\\
E-mail: {{\tt fccabarle@up.edu.ph}, {\tt hnadorna@dcs.upd.edu.ph}} \and
Research Group on Natural Computing \\
Department of Computer Science and Artificial Intelligence \\
University of Seville \\
Avda. Reina Mercedes s/n, 41012 Sevilla, Spain \\
E-mail: {\tt mdelamor@us.es}
}
\date{}

\maketitle

\begin{abstract}

We present in this paper our work regarding simulating a type of P system known as a spiking neural P system (SNP system) using graphics processing units (GPUs). GPUs, because of their architectural optimization for parallel computations, are well-suited for highly parallelizable problems. Due to the advent of general purpose GPU computing in recent years, GPUs are not limited to graphics and video processing alone, but include computationally intensive scientific and mathematical applications as well. Moreover P systems, including SNP systems, are inherently and maximally parallel computing models whose inspirations are taken from the functioning and dynamics of a living cell. In particular, SNP systems try to give a modest but formal representation of a special type of cell known as the neuron and their interactions with one another. The nature of SNP systems allowed their representation as matrices, which is a crucial step in simulating them on highly parallel devices such as GPUs. The highly parallel nature of SNP systems necessitate the use of hardware intended for parallel computations. The simulation algorithms, design considerations, and implementation are presented. Finally, simulation results, observations, and analyses using an SNP system that generates all numbers in $\mathbb N$ - \{$1$\} are discussed, as well as recommendations for future work.

\end{abstract}


\section{Introduction}

\subsection{Parallel computing: Via graphics processing units (GPUs)}
The trend for massively parallel computation is moving from the more common multi-core CPUs towards  {GPUs} for several significant reasons \cite{cudabook, cudaguide}. One important reason for such a trend in recent years include the low consumption in terms of power of  {GPUs} compared to setting up machines and infrastructure which will utilize multiple CPUs in order to obtain the same level of parallelization and performance \cite{cudapage}. Another more important reason is that  {GPUs} are architectured for \textit{massively parallel computations} since unlike most general purpose multicore CPUs, a large part of the architecture of GPUs are devoted to parallel execution of arithmetic operations, and not on control and caching just like in CPUs \cite{cudabook, cudaguide}. Arithmetic operations are at the heart of many basic operations as well as scientific computations, and these are performed with larger speedups when done in parallel as compared to performing them sequentially. In order to perform these arithmetic operations on the GPU, there is a set of techniques called $GPGPU$ (General Purpose computations on the GPU) coined by Mark Harris in 2002 which allows programmers to do computations on GPUs and not be limited to just graphics and video processing alone \cite{gpgpu}.

\subsection{Parallel computing: Via Membranes}
\textit{Membrane computing} or its more specific counterpart, a \textit{P system}, is a Turing complete computing model (for several P system variants) that perform computations nondeterministically, exhausting all possible  computations at any given time. This type of unconventional model of computation was introduced by Gheorghe P\u aun in 1998 and takes inspiration and abstraction, similar to other members of {\it Natural computing} (e.g. DNA/molecular computing, neural networks, quantum computing), from nature \cite{introtomem,ppage}. Specifically, P systems try to mimic the constitution and dynamics of the living cell: the multitude of elements inside it, and their interactions within themselves and their environment, or outside the cell's $skin$ (the cell's outermost membrane). Before proceeding, it is important to clarify what is meant when it is said that nature $computes$, particularly life or the cell: computation in this case involves reading information from memory from past or present stimuli, rewrite and retrieve this data as a stimuli from the environment, process the gathered data and act accordingly due to this processing \cite{molecular}. Thus, we try to extend the classical meaning of computation presented by Allan Turing.

SN P systems differ from other types of P systems precisely because they are $mono-membranar$ and the working alphabet contains only \textit{one object type}. These characteristics, among others, are meant to capture the workings of a special type of cell known as the $neuron$. Neurons, such as those in the human brain, communicate or 'compute' by sending indistinct signals more commonly known as action potential or \textit{spikes} \cite{snp}. $Information$ is then communicated and encoded not by the spikes themselves, since the spikes are unrecognizable from one another, but by (a) the time elapsed between spikes, as well as (b) the number of spikes sent/received from one neuron to another, oftentimes under a certain time interval \cite{snp}.

It has been shown that SN P systems, given their nature, are representable by matrices \cite{snpbrain,snpmat}. This representation allows design and implementation of an SN P system simulator using parallel devices such as GPUs.

\subsection{Simulating SNP systems in  {GPUs}}

Since the time P systems were presented, many simulators and software applications have been produced \cite{swhandbook}. In terms of \textit{High Performance Computing}, many P system simulators have been also designed for clusters of computers \cite{CiWe04}, for reconfigurable hardware as in FPGAs \cite{nguyen}, and even for GPUs \cite{amgpu,satgpu}. All of these efforts have shown that parallel architectures are well-suited in performance to simulate P systems. However, these previous works on hardware are designed to simulate \textit{cell-like} P system variants, which are among the first P system variants to have been introduced. Thus, the efficient simulation of SNP systems is a new challenge that requires novel attempts.

A matrix representation of SN P systems is quite intuitive and natural due to their graph-like configurations and properties (as will be further shown in the succeeding sections such as in subsection \ref{computesnp}). 

{On the other hand, \textit{linear algebra} operations have been efficiently implemented on parallel platforms and devices in the past years. For instance, there is a large number of algorithms implementing $matrix-matrix$ and $vector-matrix$ operations on the GPU. These algorithms offer huge performance since dense linear algebra readily maps to the data-parallel architecture of GPUs \cite{matrixgpu1,matrixgpu2}}.

It would thus seem then that a matrix represented SN P system simulator implementation on highly parallel computing devices such as  {GPUs} be a natural confluence of the earlier points made. The matrix representation of SN P systems bridges the gap between the theoretical yet still computationally powerful SN P systems and the applicative and more tangible  {GPUs}, via an SN P system simulator. 

{The design and implementation of the simulator, including the algorithms deviced, architectural considerations, are then implemented using CUDA. The Compute Unified Device Architecture (CUDA) programming model, launched by NVIDIA in mid-2007, is a hardware and software architecture for issuing and managing computations on their most recent GPU families (G80 family onward), making the GPU operate as a highly parallel computing device \cite{cudapage}. CUDA programming model extends the widely known ANSI C programming language (among other languages which can interface with CUDA), allowing programmers to easily design the code to be executed on the GPU, avoiding the use of low-level graphical primitives. CUDA also provides other benefits for the programmer such as abstracted and automated scaling of the parallel executed code.}

This paper starts out by introducing and defining the type of SNP system that will be simulated. Afterwards the NVIDIA CUDA model and architecture are discussed, baring the scalability and parallelization CUDA offers. Next, the design of the simulator, constraints and considerations, as well as the details of the algorithms used to realize the SNP system are discussed. The simulation results are presented next, as well as observations and analysis of these results. The paper ends by providing the conclusions and future work.

The objective of this work is to continue the creation of P system simulators 
, in this particular case an SN P system, using highly parallel devices such as  {GPUs}. Fidelity to the computing model (the type of SNP system in this paper) is a part of this objective.

\section{Spiking neural p systems}

\subsection{Computing with SN P systems}\label{computesnp}
The type of SNP systems focused on by this paper (scope) are those without delays i.e. those that spike or transmit signals the moment they are able to do so \cite{snpbrain,snpmat}. Variants which allow for delays before a neuron produces a spike, are also available \cite{snp}. An SNP system without delay is of the form:

\begin{definition}\label{snpdefn}
$$\Pi=(O,\sigma_1,\ldots, \sigma_m, syn, in, out),$$
where:
\begin{enumerate}
\item[1.] $O=\{a\}$ is the alphabet made up of only one object, the system spike $a$.

\item[2.] $\sigma_1,\ldots, \sigma_m$ are $m$ number of neurons of the form
$$\sigma_{i}=(n_i, R_i),1\leq i\leq m,$$
where:
\begin{enumerate}
\item[a)] $n_i\geq 0$ gives the initial number of $a$s i.e. spikes contained in neuron $\sigma_i$
\item[b)] $R_i$ is a finite set of rules of with two forms:
\begin{enumerate}
\item[(b-1)]$E/a^c \rightarrow a^p$, are known as \textit{Spiking rules}, where $E$ is a regular expression
over $a$, and $c\geq 1$, such that $p\geq 1$ number of spikes are produced, one for each adjacent neuron with $\sigma_i$ as the originating neuron and $a^c \in L(E)$.
\item[(b-2)]$a^s\rightarrow \lambda$, are known as \textit{Forgetting rules}, for $s\geq 1$, such that for each rule $E/a^c\rightarrow a$ of type (b-1) from $R_i$, $a^s\notin L(E)$. 
\item[(b-3)]$a^k \rightarrow a$, a special case of (b-1) where $E$ = $a^c$, $k \geq c$.
\end{enumerate}
\end{enumerate}
\item[3.] $syn= \{ (i,j)\, |\, 1\leq i,j \leq m, \, i\neq j\, \}$ are the synapses i.e. connection between neurons.

\item[4.] $in, out\in \{1,2,\ldots, m\}$ are the input and output neurons, respectively.
\end{enumerate}

\end{definition}

Furthermore, rules of type (b-1) are applied if $\sigma_i$ contains $k$
spikes, $a^k \in L(E)$ and $k \geq c$. Using this type of rule uses up
or consumes $k$ spikes from the neuron, producing a spike to
each of the neurons connected to it via a forward pointing arrow i.e. away from the neuron. In this manner, for rules of type (b-2)
if $\sigma_i$ contains $s$ spikes, then $s$ spikes are forgotten or
removed once the rule is used. 

\begin{figure}[h]
		\centering
		\includegraphics[scale=.35]{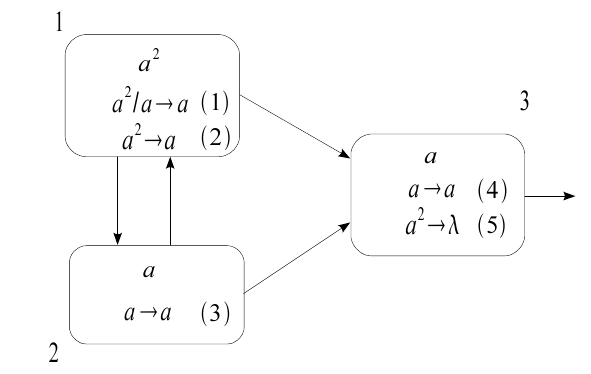} 
		\caption{An SNP P system $\Pi$, generating all numbers in $\mathbb N$ - \{$1$\}, from \cite{snpmat}.}
		\label{snp_ex}
	\end{figure}

The non-determinism of SN P systems comes with the fact
that more than one rule of the several types are applicable at a given time, given enough spikes. The rule to be used is
chosen non-deterministically in the neuron. However, only
one rule can be applied or used at a given time \cite{snp,snpbrain,snpmat}. The
neurons in an SN P system operate in parallel and in unison,
under a global clock \cite{snp}. For Figure \ref{snp_ex} no input neuron is present,
but neuron 3 is the output neuron, hence the arrow pointing
towards the environment, outside the SNP system.

The SN P system in Figure \ref{snp_ex} is $\Pi$, a 3 neuron system whose neurons are labeled (neuron 1/$\sigma_1$ to neuron 3/$\sigma_3$) and whose rules have a total system ordering from (1) to (5). Neuron 1/$\sigma_1$ can be seen to have an initial number of spikes equal to 2 (hence the $a^2$ seen inside it). There is $no$ input neuron, but $\sigma_3$ is the output neuron, as seen by the arrow pointing towards the environment (not to another neuron). More formally, $\Pi$ can be represented as follows:

$\Pi = (\{a\}, \sigma_1, \sigma_2, \sigma_3, syn, out)$ where $\sigma_1 = (2, R_1)$, $n_1 = 2, ~R_1 = \{a^2/a \rightarrow a\}$, (neurons 2 to 3 and their $n_i$s and $R_i$s can be similarly shown), $syn = \{(1,2),(1,3),(2,1),(2,3)\}$ are the synapses for $\Pi$, $out= \sigma_3$. This SN P system generates all numbers in the set $\mathbb N$ - \{$1$\}, hence it doesn't $halt$, which can be easily verified by applying the rules in $\Pi$, and checking the spikes produced by the output neuron $\sigma_3$. This generated set is the result of the computation in $\Pi$.

\subsection{Matrix representation of SNP systems}
A matrix representation of an SN P system makes use of the
following vectors and matrix definitions \cite{snpbrain,snpmat} . It is important to note that, just as in Figure \ref{snp_ex}, a total ordering of rules is {considered}.

\textit{Configuration vector} $C_k$ is the vector containing all spikes in every neuron on the $kth$ computation step/time, where $C_0$ is the initial vector containing all spikes in the system at the beginning of the computation. For $\Pi$ (in Figure \ref{snp_ex} ) the initial configuration vector is $C_0 = < 2, 1, 1 >$.

\textit{Spiking vector} shows at a given configuration $C_k$, if a
rule is applicable (has value \textit{1}) or not (has value \textit{0} instead). For $\Pi$ we have the
spiking vector $S_k = < 1, 0, 1, 1, 0 >$ given $C_0$.
Note that a 2nd spiking vector, $S_k = < 1, 0, 1, 1, 0 >$, is
possible if we use rule (2) over rule (1) instead (but not both at the same time, hence we cannot have a vector equal to
$< 1, 1, 1, 1, 0>${, so this $S_k$ is invalid }{}). $Validity$ in this case means that only one among several applicable rules is used and thus represented in the spiking vector. We can have all the possible vectors composed of $0$s and $1$s with length equal to the number of rules, but have only some of them be valid, given by $\Psi$ later at subsection \ref{siminspect}.

\textit{Spiking transition matrix} $M_{\Pi}$ is a matrix comprised of $a_{ij}$
elements where $a_{ij}$ is given as

\begin{definition}\label{defi-snp-mat}
$$
a_{ij} = \left\{
\begin{array}{rl}
-c, &\mbox{rule $r_i$ is in $\sigma_j$ and is applied consuming $c$ spikes;} \\
 p, &\mbox{rule $r_i$ is in $\sigma_s$ ($s\neq j$ and $(s,j)\in syn$)} \\
 & \mbox{and is applied producing $p$ spikes in total;}\\
 0, &\mbox{rule $r_i$ is in $\sigma_s$ ($s\neq j$ and $(s,j)\notin syn$).}
    \end{array}
\right.
$$
\end{definition}

For $\Pi$, the $M_{\Pi}$ is as follows:

\begin{equation}\label{snp_mat}
M_{\Pi} = \left(
\begin{array}{ccc}
  -1 & 1 & 1\\
  -2 & 1 &  1 \\
   1 &  -1 & 1 \\
   0 & 0 & -1\\
   0 & 0& -2
\end{array}\right)
\end{equation}

In such a scheme, rows represent rules and columns
represent neurons. 

Finally, the following equation provides the configuration
vector at the $(k+1)th$ step, given the configuration vector and
spiking vector at the $kth$ step, and $M_{\Pi}$:

\begin{equation}\label{next-config}
C_{k+1} =  C_{k} ~ + ~ S_{k}\cdot M_{\Pi}.
\end{equation}

\section{The NVIDIA CUDA architecture}

\begin{figure}
		\centering
		\includegraphics[scale=.5]{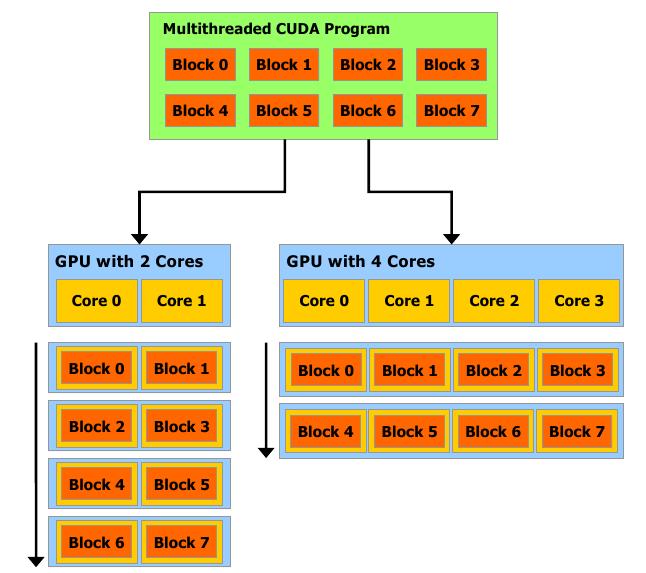} 
		\caption{NVIDIA CUDA automatic scaling, hence more cores
result to faster execution, from \cite{cudaguide}.}
		\label{cuda_scale}
	\end{figure}

NVIDIA, a well known manufacturer of GPUs, released in
2007 the CUDA programming model and architecture \cite{cudapage}.
Using extensions of the widely known C language, a
programmer can write parallel code which will then execute
in multiple threads within multiple thread blocks, each
contained within a grid of (thread) blocks. These grids
belong to a single device i.e. a single  {GPU}. Each
device/ {GPU} has multiple cores, each capable of running
its own \textit{block of threads}
The program run in the CUDA model scales
up or down, depending on the number of cores the
programmer currently has in a device. This scaling is done
in a manner that is abstracted from the user, and is
efficiently handled by the architecture as well. Automatic and efficient scaling is shown
in Figure \ref{cuda_scale}. Parallelized code will run faster with more cores
than with fewer ones \cite{cudaguide}.

	\begin{figure}
		\centering
		\includegraphics[scale=.6]{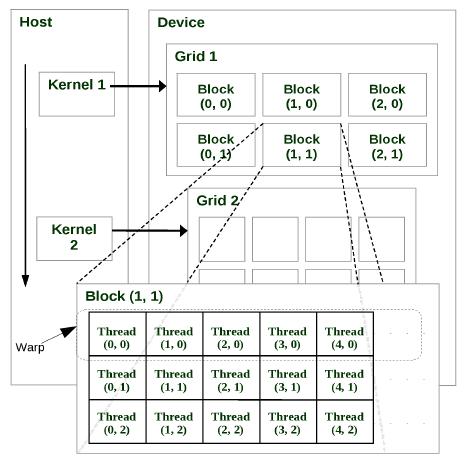} 
		\caption{NVIDIA CUDA programming model showing the sequential execution of the $host$ code alongside the parallel execution of the $kernel$ function on the $device$ side, from \cite{amgpu}.}
		\label{cuda_model}
	\end{figure}

Figure \ref{cuda_model} shows another important feature of the CUDA model:
the host and the device parts. {The host controls the execution flow while the device
is a highly-parallel co-processor.}
Device pertains to the  {GPU}/s of the system, while the host
pertains to the CPU/s. A function known as a \textit{kernel
function}, is a function called from the host but executed in
the device.

A general model for creating a CUDA enabled program is
shown in Listing \ref{cuda-code}.

\lstset{numbers=left, numberstyle=\tiny, stepnumber=1, numbersep=5pt, label=cuda_gencode}

\begin{lstlisting}[label=cuda-code,caption=General code flow for CUDA programming written in the CUDA extended C language]
//allocate memory on GPU e.g.
cudaMalloc( (void**)&dev_a, N * sizeof(int)

//populate arrays
 . . .

//copy arrays from host to device e.g.
cudaMemcpy( dev_a, a, N * sizeof(int), 
cudaMemcpyHostToDevice)

//call kernel (GPU) function e.g.
add<<<N, 1>>>( dev_a, dev_b, dev_c );

// copy arrays from device to host e.g.
cudaMemcpy( c, dev_c, N * sizeof( int), 
cudaMemcpyDeviceToHost )

//display results

//free memory e.g.
cudaFree( dev_a );
\end{lstlisting}

Lines 2 and 21, implement CUDA versions of
the standard C language functions e.g. the standard C
function $malloc$ has the CUDA C function counterpart being
$cudaMalloc$, and the standard C function $free$ has $cudaFree$ as its CUDA C counterpart.

Lines 8 and 15 show a CUDA C specific function, namely
$cudaMemcpy$, which, given an input of pointers ( from Listing \ref{cuda-code} host code
pointers are single letter variables such as $a$ and $c$,while
device code variable counterparts are prefixed by $dev\_$ such
as $dev\_a$ and $dev\_c$ ) and the size to copy ( as computed by
the $sizeof$ function ), moves data from host to device
( parameter $cudaMemcpyHostToDevice$ ) or device to host
( parameter $cudaMemcpyDeviceToHost$).

A kernel function call uses the {triple $<$ and $>$} operator, in
this case the kernel function 

{$add< < < N, 1 > > >$}$( dev\_a, dev\_b,
dev\_c )$.

This function adds the values, per element (and
each element is associated to 1 thread), of the variables
$dev\_a$ and $dev\_b$ sent to the device, collected in variable
$dev\_c$ before being sent back to the host/CPU. The variable
$N$ in this case allows the programmer to specify $N$ number of
threads which will execute the $add$ kernel function in parallel, with 1
specifying only one block of thread for all $N$ threads.

\subsection{Design considerations for the hardware and software
setup}

Since the kernel function is executed in parallel in
the device, the function needs to have its {inputs}
initially moved from the CPU/host to the device, and then back from the device to the
host after computation { for the results}. This movement of data back and forth should be minimized in order to obtain more efficient,
in terms of time, execution.
Implementing an equation such as (\ref{next-config}), which involves
multiplication and addition between vectors and a matrix,
can be done in parallel with the previous considerations in
mind. In this case, $C_k$, $S_k$, and $M_{\Pi}$ are loaded, manipulated, and pre-processed within the host code,
before being sent to the kernel function which will perform
computations on these function arguments in parallel.
To represent $C_k$, $S_k$, and $M_{\Pi}$, text files are created 
to house each input, whereby each element of the
vector or matrix is entered in the file in order, from left to right, with a blank space in between as a delimiter. The
matrix however is entered in row-major ( a linear array of all
the elements, rows first, then columns) order format i.e. for
the matrix $M_{\Pi}$ seen in (\ref{snp_mat}), the row-major order version is simply
\begin{equation}\label{row-maj}
-1, 1, 1, -2, 1, 1, 1, -1, 1, 0, 0, -1, 0, 0, -2
\end{equation}

Row major ordering is a well-known ordering and
representation of matrices for their linear as well as parallel
manipulation in corresponding algorithms \cite{cudabook}. Once all
computations are done for the $(k+1)th$ configuration, the result of
equation (\ref{next-config}) are then collected and moved from the
device back to the host, where they can once again be
operated on by the host/CPU. It is also important to note
that these operations in the host/CPU provide logic and
control of the data/inputs, while the device/GPU provides
the arithmetic or computational 'muscle', the laborious task
of working on multiple data at a given time in parallel,
hence the current dichotomy of the CUDA programming model \cite{amgpu}. The GPU acts as a \textit{co-processor} of the central processor.
This division of labor is observed in Listing \ref{cuda-code} .

\subsection{Matrix computations and CPU-{GPU} interactions}

Once all 3 initial and necessary inputs are loaded, as is to be
expected from equation \ref{next-config}, the device is first instructed to
perform multiplication between the spiking vector $S_k$ and the
matrix $M_{\Pi}$. To further simplify computations at this point, the
vectors are treated and automatically formatted by the host
code to appear as single row matrices, since vectors can be
considered as such. Multiplication is done per element (one
element is in one thread of the device/GPU{}), and then the
products are collected and summed to produce a single
element of the resulting vector/single row matrix.

Once multiplication of the $S_k$ and $M_{\Pi}$ is done,
the result is added to the $C_k$, once
again element per element, with each element belonging to
one thread, executed at the same time as the others.

For this simulator, the host code consists largely of
the programming language \textit{Python}, a well-known high-
level, object oriented programming (OOP) language. The
reason for using a high-level language such as Python is
because the initial inputs, as well as succeeding ones
resulting from exhaustively applying the rules and equation
(\ref{next-config}) require manipulation of the vector/matrix elements or
values as \textit{strings}. The strings are then concatenated, checked on (if they
conform to the form (b-3) for example) by the host, as well
as manipulated in ways which will be elaborated in the
following sections along with the discussion of the
algorithm for producing all possible and valid $S_k$s and $C_k$s given initial conditions. The host code/Python part thus implements the logic and control as mentioned earlier, while in it, the device/GPU code which is written in C executes the parallel parts of the simulator for CUDA to be utilized.

\section{Simulator design and implementation}\label{sect-snp-algo}

The current SNP simulator, which is based on the type of
SNP systems without time delays, is capable of
implementing rules of the form (b-3) i.e. whenever the
regular expression $E$ is equivalent to the regular expression $a^k$ in that rule. Rules are entered in the same manner
as the earlier mentioned vectors and matrix, as blank space
delimited values (from one rule to the other, belonging to the same neuron) and \$ delimited ( from one neuron to the
other). Thus for the SNP system ${\Pi}$ shown earlier, the file $r$
containing the blank space and \textit{\$} delimited values is as
follows:
\begin{equation}\label{rules}
2~2~\$~1~\$~1~2
\end{equation}

That is, rule (1) from Figure \ref{snp_ex} has the value $2$ in the file $r$ (though rule (1)
isn't of the form (b-3) it nevertheless consumes a spike since
its regular expression is of the same regular expression type
as the rest of the rules of ${\Pi}$ ). Another implementation
consideration was the use of $lists$ in Python, since unlike
dictionaries or tuples, lists in Python are  \textit{mutable}, which is a
direct requirement of the vector/matrix element
manipulation to be performed later on (concatenation
mostly). Hence a $C_k = <2, 1, 1>$ is
represented as $[ 2, 1, 1 ]$ in Python. That is, at the $kth$
configuration of the system, the number of spikes of neuron
1 are given by accessing the index (starting at zero) of the
configuration vector Python $list$ variable $confVec$, in this case if
\begin{equation}\label{confvec}
confVec = [ 2, 1, 1 ]
\end{equation}
	
then $confVec[ 0 ] = 2$ gives the number of spikes available at
that time for neuron 1, $confVec[ 1 ] = 1$ for neuron 2, and so
on. The file $r$, which contains the ordered list of neurons and
the rules that comprise each of them, is represented as a \textit{list of sub-
lists} in the Python/host code. For SNP system ${\Pi}$ and from (\ref{rules}) we have the
following:
\begin{equation}\label{rule-list}
r = [ [ 2, 2 ], [ 1 ], [ 1, 2 ] ]
\end{equation}

Neuron 1's rules are given by accessing the sub-lists of $r$
(again, starting at index zero) i.e. rule (1) is given by $r[ 0 ][ 0 ]
= 2$ and rule (4) is given by $r[ 2 ][ 1 ] = 1$.
Finally, we have the input file $M$, which holds the Python $list$ version of (\ref{row-maj}).

\subsection{Simulation algorithm implementation}\label{snp-sim-algo}

The general algorithm is shown in Algorithm \ref{sim-algo}. {Each line in Algorithm \ref{sim-algo} mentions
which part/s the simulator code runs in, either in the device \textbf{(DEVICE)} or in the host \textbf{(HOST)} part.} 
Step $IV$ of Algorithm \ref{sim-algo} makes
the algorithm stop with \textit{2 stopping criteria} to do this:

One is when there are no more available spikes in the system (hence a
zero value for a configuration vector), and the second one
being the fact that all previously generated configuration
vectors have been produced in an earlier time or
computation, hence using them again in part I of Algorithm \ref{sim-algo} would be pointless, since a
redundant, infinite loop will only be formed.

\begin{algorithm}                   
\caption{Overview of the algorithm for the SNP system simulator}         \label{sim-algo}                           
\begin{algorithmic}                    
\REQUIRE Input files: $confVec$, $M,r$.

I. \textbf{(HOST)} Load input files.
Note that $M$ and $r$ need only be loaded once since they are unchanging, $C_0$ is loaded once, and then $C_k$s are loaded afterwards.

II. \textbf{(HOST)} Determine if a rule/element in $r$ 
is applicable based on its corresponding spike value 
in $confVec$, and then generate all valid and
possible spiking vectors in a list of lists $spikVec$ given the 3 initial inputs.

III. \textbf{(DEVICE)} From part II., run the kernel 
function on $spikVec$, which contains all 
the valid and possible spiking vectors for 
the current $confVec$ and $r$. This will 
generate the succeeding $C_k$s and their corresponding $S_k$s.

IV. \textbf{(HOST+DEVICE)} Repeat steps I to IV (except instead of loading $C_0$ as $confVec$, use the generated $C_k$s in III) until
a zero configuration vector (vector with 
only zeros as elements) or further $C_k$s
produced are repetitions of a $C_k$
produced at an earlier time. (Stopping criteria in subsection \ref{snp-sim-algo} )

\end{algorithmic}
\end{algorithm}	

Another
important point to notice is that either of the stopping
criterion from \ref{snp-sim-algo} could allow for a deeply nested computation tree,
one that can continue executing for a significantly lengthy amount of
time even with a multi-core CPU and even the more parallelized
 {GPU}.

\subsection{Closer inspection of the SN P system
simulator}\label{siminspect}
	
The more detailed algorithm for part $II$ of Algorithm \ref{sim-algo} is as follows.

Recall from the definition of an SNP system (Definitin \ref{snpdefn}) that we have $m$ number of $\sigma$s. We related $m$ to our implementation by noticing the cardinality of the Python list $r$.
\begin{equation}\label{rcard}
|r| = m
\end{equation}

{
\begin{equation}\label{psi}
\Psi = |{\sigma_{V_1}}| |{\sigma_{V_2}}| ...|{\sigma_{V_m}}|
\end{equation}
}

where

$|{\sigma_{V_m}}|$

means the total number of rules in the $mth$ neuron which satisfy the regular expresion $E$ in (b-3). $m$ gives the total number of neurons, while $\Psi$ gives the expected number of $valid$ and $possible$ $S_k$s which should be produced in a given configuration.
We also define $\omega$ as both the largest and last integer value in the sub-list (neuron) created in step II of Algorithm \ref{sim-algo} and further detailed in Algorithm \ref{sim-algo2}, which tells us how many elements of that neuron satisfy $E$.

During the exposition of the algorithm, the previous Python
lists (from their vector/matrix counterparts in earlier
sections) (\ref{confvec}) and (\ref{rule-list}) will be utilized. For part $II$ Algorithm \ref{sim-algo} we have a sub-algorithm (Algorithm \ref{sim-algo2}) for generating all valid
and possible spiking vectors given input files $M$, $confVec$, and $r$.

\begin{algorithm}
\caption{Algorithm further detailing part II in Algorithm \ref{sim-algo}}
\label{sim-algo2}
\begin{algorithmic}
\item [II-1.] Create a list $tmp$, a copy of $r$, marking
each element of $tmp$ in increasing order of $\mathbb N$,
as long as the element/s satisfy the rule's 
regular expression $E$ of a rule (given by list
$r$ ). Elements that don't satisfy $E$ are marked with 0.\\
 
\item [II-2.] To generate all possible and valid spiking vectors from
$tmp$, we go through each neuron i.e. all elements of $tmp$,
since we know a priori $m$ as well as the number of
elements per neuron which satisfy $E$. We only need to iterate
through each neuron/element of $tmp$, $\omega$ times. 
(from II-1). We then produce a new list, $tmp2$, which is
made up of a sub-list of strings from all possible and valid
\textit{\{1,0\}} strings i.e. spiking vectors per neuron.\\

\item  [II-3.] To obtain all possible and valid \textit{\{1,0\}}
strings ($S_k$s), given that there are multiple strings
to be concatenated ( as in $tmp2$'s case ), pairing up the
neurons first, in order, and then exhaustively distributing
every element of the first neuron to the elements of the 2nd
one in the pair. These paired-distributed strings will be stored in a new list, $tmp3$.

\end{algorithmic}
\end{algorithm}

Algorithm \ref{sim-algo2} ends once all \textit{\{1,0\}} have been paired up to one another. As an illustration of Algorithm \ref{sim-algo2}, consider (\ref{confvec}), (\ref{rule-list}), and (\ref{snp_mat}) as inputs to our SNP system simulator. The following details the production of all valid and possible spiking vectors using Algorithm \ref{sim-algo2}.

Initially from II-1 of Algorithm \ref{sim-algo2}, we have

$r = tmp = [ [ 2, 2 ], [ 1 ], [ 1, 2 ] ]$.

Proceeding to illustrate II-2 we have the following passes.

1st pass:
$tmp = [ [ 1, 2 ], [ 1 ], [ 1, 2 ] ]$\\
\textit{Remark/s}: previously, $tmp [ 0 ][ 0 ]$ was equal to 2,
but now has been changed to 1, since it satisfies $E$
( $configVec[ 0 ] = 2$ w/c is equal to 2, the
number of spikes consumed by that rule).$\Sigma$

2nd pass:
$tmp = [ [ 1, 2 ], [ 1 ], [ 1, 2 ] ]$\\
\textit{Remark/s}: previously $tmp[ 0 ][ 1 ] = 2$, which has
now been changed (incidentally) to 2 as well, since
it's the 2nd element of $\sigma_1$ which satisfies $E$.

3rd pass:
$tmp = [ [ 1, 2 ], [ 1 ], [ 1, 2 ] ]$\\
\textit{Remark/s}: 1st (and only) element of neuron 2 which
satisfies $E$.

4th pass:
$tmp = [ [ 1, 2 ], [ 1 ], [ 1, 2 ] ]$\\
\textit{Remark/s}: Same as the 1st pass

5th pass:
$tmp = [ [ 1, 2 ], [ 1 ], [ 1, 0 ] ]$\\
\textit{Remark/s}: element $tmp[ 2 ][ 1 ]$, or the 2nd
element/rule of neuron 3 doesn't satisfy $E$.

Final result:
$tmp = [ [ 1, 2 ], [ 1 ], [ 1, 0 ] ]$

At this point we have the following, based on the
earlier definitions:

$m$ = 3 ( 3 neurons in total, one per element/value of
$confVec$)

$\Psi = |\sigma_{V_1}| |\sigma_{V_2}| |\sigma_{V_3}| = 2 * 1 * 1 = 2$

$\Psi$ tells us the number of valid strings of \textit{1}s and \textit{0}s i.e.
$S_k$s, which need to be produced later, for a
given $C_k$ which in this case is $confvec$. There are only
2 valid $S_k$s/spiking vectors from (\ref{confvec}) and the
rules given in (\ref{rule-list}) encoded in the Python list $r$. These $S_k$s are
\begin{equation}\label{sk-211-1}
< 0, 1, 1, 1, 0>
\end{equation}
\begin{equation}\label{sk-211-2}
< 1, 0, 1, 1, 0>
\end{equation}
In order to produce all $S_k$s in an algorithmic way as is done in Algorithm \ref{sim-algo2} , it's
important to notice that first, \textit{all possible and valid} $S_k$s ( made up of \textit{1}s and \textit{0}s) per $\sigma$ have to be
produced first, which is facilitated by II-1 of Algorithm \ref{sim-algo2} and its output (the
current value of the list $tmp$ ).

Continuing the illustration of II-1, and illustrating II-2 this time, we iterate over neuron 1 twice, since its $\omega$ = 2, i.e. neuron 1 has only 2 elements which satisfy $E$, and consequently, it is its 2nd element,

		$tmp[ 0 ] [ 1 ] = 2.$ 

For neuron 1, our first pass along its elements/list is as follows. Its 1st element,

		$tmp[ 0 ][ 0 ] = 1$

is the first element to satisfy $E$, hence it requires a \textit{1} in its place, and \textit{0} in the others. We therefore produce the string	'\textit{10}' for it. Next, the 2nd element satisfies $E$ and it too, deserves a \textit{1}, while the rest get \textit{0}s. We produce the string '\textit{01}' for it. 

The new list, $tmp2$, collecting the strings produced for neuron 1 therefore becomes

		$tmp2 = [ [ 10, 01 ] ] $

Following these procedures, for neuron 2 we get $tmp2$ to be as follows: 

		$tmp2 = [ [ 10, 01 ], [ 1 ] ]$ 

Since neuron 2 which has only one element only has 1 possible and valid string, the string 1. 
Finally, for neuron 3, we get $tmp2$ to be 

		$tmp2 = [ [ 10, 01 ], [ 1 ], [ 10 ] ] $

In neuron 3, we iterated over it only once because $\omega$, the number of elements it has which satisfy $E$, is equal to 1 only. 
Observe that the sublist 

		$tmp2[ 0 ] = [ 10, 01 ]$ 

is equal to all possible and valid \{\textit{1,0}\} strings for neuron 1, given rules in (\ref{rule-list}) and the number of spikes in $configVec$. 

Illustrating II-3 of Algorithm \ref{sim-algo2}, given the valid and
possible \{\textit{1,0}\} strings (spiking vectors) for neurons 1, 2, and 3 (separated per neuron-column) from (\ref{confvec}) and (\ref{rule-list}) and from the illustration of II-2, all possible and valid list of \{\textit{1,0}\} string/s for neuron 1: ['10','01'], neuron 2: ['1'], and neuron 3: ['10'].

First, pair the strings of neurons 1 and 2, and then distribute them exhaustively to the other neuron's possible and valid strings, concatenating them in the process (since they are considered as $strings$ in Python). 

'10'  +	'1' $\rightarrow$ '101'

'01' 
\\
and 

'10'

'01'  +	'1' $\rightarrow$ '011'

now we have to create a new list from $tmp2$, which will house the concatenations we'll be doing. In this case, 

		$tmp3 = [ 101, 011 ] $

next, we pair up $tmp3$ and the possible and valid strings of neuron 3 

'101' + '10' $\rightarrow$ '10110'

'011'
\\ 
and 

'101'

'011' + '10' $\rightarrow$ '01110' 

eventually turning $tmp3$ into 

		$tmp3 = [ 10110, 01110 ]$ 

The final output of the sub-algorithm for the generation of all valid and possible spiking vectors is a list, 

		$tmp3 = [ 10110, 01110 ] $

As mentioned earlier, $\Psi$ = 2 is the number of valid and possible $S_k$s to be expected from $r$, $M_\Pi$, and $C_0$ = [2,1,1] in $\Pi$. Thus $tmp3$ is the list of all possible and valid spiking vectors given (\ref{confvec}) and (\ref{rule-list}) in this illustration. Furthermore, $tmp3$ includes all possible and valid spiking vectors for a given neuron in a given configuration of an SN P system with all its rules and 
synapses (interconnections). Part II-3 is done ( $m - 1$) times, albeit exhaustively still so, between the two lists/neurons in the pair.

\section{Simulation results, observations, and analyses}
The SNP system simulator (combination of Python
and CUDA C) implements the algorithms in section \ref{sect-snp-algo}
earlier. A sample simulation run with the SNP system $\Pi$ is
shown below (most of the output has been truncated due to space constraints ) with $C_0$ = [2,1,1]
\begin{verbatim}
****SN P system simulation run STARTS here****
Spiking transition Matrix: 
... 

Rules of the form a^n/a^m -> a or a^n ->a loaded: 
['2', '2', '$', '1', '$', '1', '2'] 

Initial configuration vector: 211
 
Number of neurons for the SN P system is 3 
Neuron 1  rules criterion/criteria and total order 
... 

 tmpList =  [['10', '01'], ['1'], ['10']] 
 All valid spiking vectors: allValidSpikVec =
 [['10110', '01110']] 
All generated Cks are allGenCk =
['2-1-1', '2-1-2', '1-1-2'] 
 End of C0 
** 
** 
** 
 initial total Ck list is
   ['2-1-1', '2-1-2', '1-1-2'] 
Current confVec: 212 
All generated Cks are allGenCk =
['2-1-1', '2-1-2', '1-1-2', '2-1-3', '1-1-3'] 
** 
** 
** 
Current confVec: 112 
All generated Cks are allGenCk =
['2-1-1', '2-1-2', '1-1-2', '2-1-3', '1-1-3',
'2-0-2', '2-0-1'] 
** 
** 
...

Current confVec: 109
All generated Cks are allGenCk = ['2-1-1', '2-1-2',
...
 '1-0-7', '0-1-9', '1-0-8', '1-0-9']

**
**
**

No more Cks to use (infinite loop/s otherwise). Stop.
****SN P system simulation run ENDS here****
\end{verbatim}

That is, the computation tree for SNP system $\Pi$ with $C_0$ = [2,1,1] went down as deep as $confVec = 109$. At that point, all configuration vectors for all possible and valid spiking vectors have been produced. The Python list variable $allGenCk$ collects all the $C_k$s produced. In Algorithm \ref{sim-algo2} all the values of $tmp3$ are added to $allGenCk$. The final value of $allGenCk$ for the above simulation run is

~\\
\textit{
allGenCk = ['2-1-1', '2-1-2', '1-1-2', '2-1-3', '1-1-3', '2-0-2', '2-0-1', '2-1-4', '1-1-4', '2-0-3', '1-1-1', '0-1-2', '0-1-1', '2-1-5', '1-1-5', '2-0-4', '0-1-3', '1-0-2', '1-0-1', '2-1-6', '1-1-6', '2-0-5', '0-1-4', '1-0-3', '1-0-0', '2-1-7', '1-1-7', '2-0-6', '0-1-5', '1-0-4', '2-1-8', '1-1-8', '2-0-7', '0-1-6', '1-0-5', '2-1-9', '1-1-9', '2-0-8', '0-1-7', '1-0-6', '2-1-10', '1-1-10', '2-0-9', '0-1-8', '1-0-7', '0-1-9', '1-0-8', '1-0-9'] }

\begin{figure}[tb]
		\centering
		\includegraphics[scale=.7]{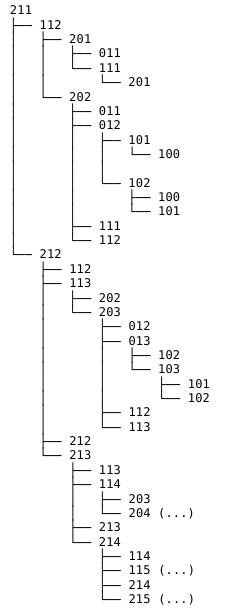} 
		\caption{The computation tree graphically representing the output of the simulator run over $\Pi$ with $C_0$ = [2, 1, 1]}
		\label{c211_tree}
	\end{figure}

~\\
 It's also noteworthy that the simulation for $\Pi$ didn't stop at the 1st stopping criteria (arriving at a zero vector i.e. $C_k$ = [0,0,0] ) since $\Pi$ generates  all natural counting numbers greater than 1, hence a loop (an infinite one) is to be expected. The simulation run shown above stopped with the 2nd stopping criteria from Section \ref{sect-snp-algo}. Thus the simulation was able to exhaust all possible configuration vectors and their spiking vectors, stopping only since a repetition of an earlier generated $confVec$/$C_k$ would introduce a loop (triggering the 2nd stopping criteria in subsection \ref{snp-sim-algo}).
Graphically (though not shown exhaustively) the computation tree for $\Pi$ is shown in Figure \ref{c211_tree}.

The $confVecs$ followed by (...) are the $confVecs$ that went deeper i.e. produced more $C_k$s than Figure \ref{c211_tree} has shown.

\section{Conclusions and future work}
Using a highly parallel computing device such as a  {GPU}, {and the}
 NVIDIA CUDA {programming model}, an SNP system simulator was
successfully designed and implemented as per the objective of this work. The simulator was shown to model the workings of an SN P system without delay using the system's matrix representation. The use of a high
level programming language such as Python for host tasks,
mainly for logic and string representation and manipulation
of values (vector/matrix elements) {has} provided the necessary
expressivity to implement the algorithms created to produce
and exhaust all possible and valid configuration and spiking
vectors. For the device tasks, CUDA allowed the
manipulation of the NVIDIA CUDA enabled  {GPU} which
took care of repetitive and highly parallel computations
({vector-matrix} addition and multiplication essentially).

Future versions of the SNP system simulator will focus on
several improvements. These improvements include the use
of an {optimized} algorithm for matrix computations {on the GPU} without requiring
the input matrix to be transformed into a square matrix (this is
currently handled by the simulator by padding zeros to an
otherwise non-square matrix input). Another improvement
would be the simulation of systems not of the form (b-3).
Byte-compiling the Python/host part of the code to improve
performance as well as metrics to further enhance and
measure execution time are desirable as well. Finally, deeper
understanding of the CUDA architecture, such as inter-
thread/block communication, for very large systems
with equally large matrices, is required. These
improvements as well as the current version of the simulator
should also be run in a machine or setup with higher versions of
 {GPUs} {supporting} NVIDIA CUDA.

\section{ Acknowledgments }
Francis Cabarle is supported by the {DOST-ERDT scholarship program}. Henry Adorna is funded by the {DOST-ERDT research grant} and the Alexan professorial chair of the {UP Diliman Department of Computer Science, University of the Philippines Diliman}. They would also like to acknowledge the {Algorithms and Complexity laboratory} for the use of Apple iMacs with NVIDIA CUDA enabled GPUs for this work. {Miguel A. Mart\'inez--del--Amor is supported by ``Proyecto de Excelencia con Investigador de Reconocida Val\'ia'' of the ``Junta de Andaluc\'{\i}a'' under grant P08-TIC04200, and the support of the project TIN2009--13192 of the ``Ministerio de Educaci\'on y Ciencia'' of Spain, both co-financed by FEDER funds.}
Finally, they would also like to thank the valuable insights of Mr. Neil Ibo.

\end{document}